\documentclass[seceq]{ptptex}

\usepackage{graphicx}

\usepackage{wrapft}

\title{
On the origin of nuclear clustering
}

\author{
Jacek \textsc{Oko{\l}owicz}$^{1,}$\footnote{E-mail: Jacek.Okolowicz@ifj.edu.pl}, 
Marek \textsc{P{\l}oszajczak}$^{2,}$\footnote{E-mail: ploszajczak@ganil.fr}
and Witold \textsc{Nazarewicz}$^{3-5,}$\footnote{E-mail: witek@utk.edu} 
}
\inst{
$^1$Institute of Nuclear Physics, Radzikowskiego 152, PL-31342 Krak\'ow, Poland\\
$^2$Grand Acc\'el\'erateur National d'Ions Lourds (GANIL), CEA/DSM - CNRS/IN2P3,
BP 55027, F-14076 Caen Cedex, France\\
$^3$Department of Physics and Astronomy, University of Tennessee, Knoxville, Tennessee 37996, USA\\
$^4$Physics Division, Oak Ridge National Laboratory, Oak Ridge, Tennessee 37831, USA\\
$^5$Institute of Theoretical Physics, University of Warsaw, ul. Ho\.za 69,
PL-00-681 Warsaw, Poland
}

\abst{
Clustering is one of the most complex  phenomena known to the structure of atomic nuclei. A comprehensive description of this ubiquitous phenomenon  goes beyond standard shell model and cluster model frameworks. We argue that  clustering  is a consequence of an openness of the nuclear many-body system. To illustrate this point, we  study the near-threshold behavior of exceptional points  in $^{16}$Ne and $^{24}$S.}

\begin{document}
\maketitle

\section{Introduction}
\label{sec1}
The nuclear many-body problem is among the most difficult challenges in physics due to its paramount complexity and importance at various scales: from  sub-nucleon to astronomical. This  complexity is  the reason that the theory of the atomic nucleus is still very much a ``work in progress.'' A comprehensive, unified description of all nuclei at the nucleonic level would require merging structural approaches like ab initio, configuration-interaction (shell model), and nuclear density functional theory with reaction  approaches such as the continuum shell model or the modern theory of optical potential. In this quest, we need new theoretical concepts and new experimental data, especially close to the drip lines where non-local effects appear already in the ground state.

Since such a coherent description is not yet within our grasp, however, one avenue is to create a patchwork of advanced  models describing  selected nuclear features. Consequently, our understanding of nuclear structure is incomplete and even incoherent in certain aspects.  Low-energy excitations in light atomic nuclei provide the case in point. Both bound states and resonances can be   modeled using  the nuclear shell model (SM) and  the cluster model (CM), but the picture of how nucleons are organized within  many-body states is different in these approaches. The 
degrees of freedom  of SM are the valence protons and neutrons moving in selected shells\cite{rf:1}, whereas the CM is based on a molecular viewpoint that employs more composite effective building blocks\cite{rf:2} such as $\alpha-$particles. Moreover, both models  assume the nucleus to be a closed quantum system (CQS) that is completely isolated from the space of  scattering and decay channels.  Can the comprehensive understanding of low-energy excitations possibly emerge from such disjointed physical pictures?

In this paper, we shall highlight some of the theoretical issues underlying the clustering phenomenon by analyzing properties of the continuum shell model (CSM) wave functions\cite{rf:3,rf:4} in the neighborhood of the proton decay threshold. Salient features of near-threshold CSM states will be illustrated by studying exceptional points\cite{rf:5} (EPs) of the complex-extended CSM Hamiltonian. Exceptional points are singularities of eigenvalues and eigenvectors for complex values of selected model parameters; they can be associated with  square root branch point singularities of the eigenvalues in the vicinity of level repulsion. By investigating the energy- and (complex) interaction-dependence of exceptional points around the proton emission threshold and  identifying  salient features of eigenfunctions due to a coupling to a common decay threshold, we hope to reveal generic features of the clusterization mechanism.

Our paper is organized as follows. In Sec. \ref{sec2} we  examine physical arguments that necessitate the use of an OQS description of the  clustering phenomenon and reiterate its universal features. A brief summary of the real-energy CSM and EPs is  presented in Sec. \ref{sec3}. The properties of exceptional threads, i.e., trajectories of EPs  in the space of system energy and continuum-coupling interaction, is discussed in Sec. \ref{sec4}. As an illustrative example, we  study  the $J^{\pi}=0^+$ CSM eigenfunctions of $^{16}$Ne and $^{24}$S, which are mixed due to the coupling to a common $\ell=0$ proton decay channel. These results provide a good insight into the  configuration mixing mechanism operating in more complicated decay channels involving composite charged particles such as  $^3$H, $^2$He, $^4$He, and $^8$Be. Finally, Sec. \ref{sec5}
contains the main conclusions of this work.

\section{Salient configuration mixing phenomena near the dissociation threshold}
\label{sec2}

What can be said about the structure of many-body states  in the narrow range of energies around the reaction threshold? Are those properties very dependent on a  particular realization of the Hamiltonian? Is there a connection between the branch point singularity at the particle emission threshold and the appearance of cluster states? 
In the context of these questions,  the observation by Ikeda et al. \cite{rf:6} offers  
important insight:  $\alpha$-cluster states can be found in the proximity of  $\alpha$-particle decay thresholds. Curiously, this observation has not raised much interest in studies of the coupling to particle-decay channels as the mechanism behind enhanced correlations in near-threshold SM states. 

The conjecture of Ikeda et al. \cite{rf:6} can be formulated more generally: the coupling to a nearby particle/cluster decay channel induces particle/cluster correlations in SM wave functions which constitute the imprint of this channel. In other words, the clustering is the generic near-threshold phenomenon in OQS that does not originate from any particular property of the Hamiltonian or some symmetry of the nuclear many-body problem. 
\begin{figure}[htb]
\begin{center}
\includegraphics[scale=0.6]{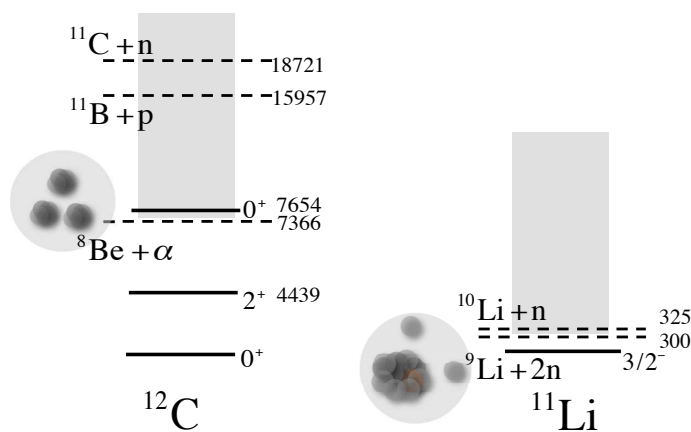}
\caption{\label{fig1}Example of two Borromean systems: $^{12}$C and $^{11}$Li. The near-threshold states in these nuclei -- the Hoyle resonance $0^+_2$ in $^{12}$C  and the weakly-bound $3/2^-_1$ ground state in $^{11}$Li --  exhibit a strong imprint of the nearby decay threshold, i.e.,   $^{12}$C$\rightarrow ^8$Be+$\alpha$ and $^{11}$Li$\rightarrow 2n+^9$Li.}
\end{center} 
\end{figure}
We claim that this conjecture holds for all kinds of cluster states including unstable systems like dineutron, or $^8$Be. Figure \ref{fig1} shows two examples of the Borromean systems for which the lowest-energy threshold corresponds to the emission of a cluster resonance: $^8$Be in $^{12}$C and dineutron in $^{11}$Li. 

Universality of the clustering phenomenon stems from basic properties of the scattering matrix in a multichannel system\cite{rf:7}. Indeed, the decay threshold is a branching point of the scattering matrix. For energies below the lowest particle-decay threshold at energy $E_1$, one finds the analytic phase with a single solution which is regular in the entire space. Above $E_1$ and below the energy $E_2$ of the  next decay channel, the new analytic phase corresponds to two regular solutions of the scattering problem, etc. Hence, one obtains a set of analytic phases, each one with a different number of regular scattering solutions. These phases are separated by decay thresholds and, together, form the multichannel coupled OQS.
\begin{figure}[htb]
\begin{center}
\includegraphics[scale=0.45]{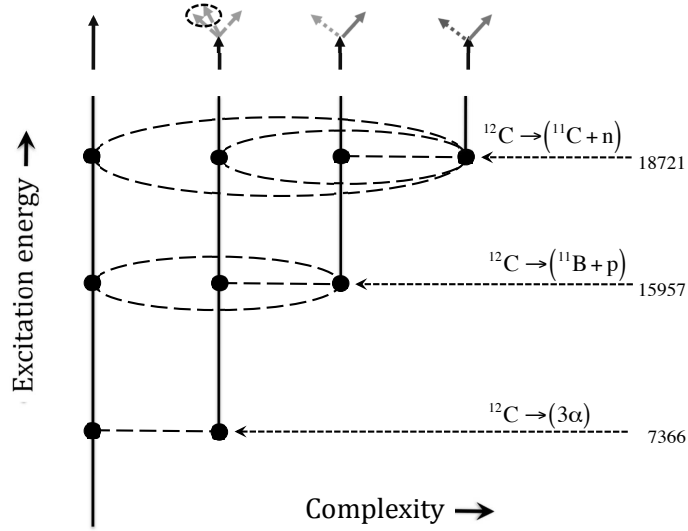}
\caption{The schematic picture of couplings in the multichannel representation of $^{12}$C. With the increasing excitation energy, subsequent decay channels open up at threshold energies $E_{1}$, $E_{2}$, $E_{3}$, $\dots$, leading to a complex multichannel network of couplings. When a new channel opens up at the threshold  $E_i$, the unitarity imposes the appearance of new channel couplings; hence,  a modification of all eigenfunctions with energies $E<E_i$.}
\label{fig2}
\end{center}
\end{figure}

Each decay threshold (branching point) can be associated with  a non-analytic point of the scattering matrix. The coupling of different SM eigenfunctions to the same decay channel induces a mixing among them which reflects the nature of the branching point. In this way, regular scattering solutions are modified in the proximity of each threshold. 
Such configuration mixing involving {\em all}  SM eigenfunction of the same quantum numbers (angular momentum and parity) can radically change the structure of near-threshold SM states. This is the reason for coexistence of `good SM' and `good CM' states in the spectra of light nuclei. That is also why SM and CM  schemes can be  employed to model wave functions in analytic phases.

The flux conservation (unitarity) in a multichannel system implies that the mixing of SM eigenfunctions changes whenever  a new channel opens up. In this sense, the eigenfunctions are not immutable characteristics of the system but they strongly depend  on the  energy window imposed theoretically and vary with the increasing total energy of the system. The role of unitarity and channel coupling  for near-threshold eigenfunctions has been demonstrated within an OQS formalism  for  spectroscopic factors\cite{rf:8} and radial overlap integrals\cite{rf:9}.

Figure \ref{fig2} illustrates the multichannel network of couplings on the example of  $^{12}$C. For the excitation energies $E$ below the first decay threshold $E_{^{12}{\text C}\rightarrow 3\alpha}$, the regular phase contains only elastic scattering solutions. For $E>E_{^{12}{\text C}\rightarrow 3\alpha}$ and below the second decay threshold at  $E_{^{12}{\text C}\rightarrow ^{11}B+p}$, the regular phase contains two solutions coupled by the unitarity condition, and the story goes on  at higher excitation energies. The structure of any eigenfunction in this multichannel OQS depends on the total energy of the system, which in turn determines the environment of decay channels accessible for couplings. With increasing excitation energy, the multichannel OQS network of couplings becomes increasingly  more complex.

\section{Open quantum system formulation of the nuclear shell model}
\label{sec3}

In the standard CQS formulation of the nuclear  SM, nucleons occupy bound single particle orbits of an infinite (harmonic oscillator) potential; hence, they are isolated from the environment of scattering states\cite{rf:1}. Since the scattering continuum is not considered, the presence of decay thresholds and exceptional points is  neglected. Hence, the essential aspect of the nuclear clustering mechanism is totally absent. In  CM, couplings to decay channels are put by hand by  introducing {\it a posteriori} suitable effective cluster degrees of freedom. 

To formulate the nuclear  SM for OQSs, two schemes have been proposed. The first one, the Gamow Shell Model (GSM)\cite{rf:10}, is the complex-energy CSM based on  the one-body Berggren ensemble\cite{rf:11}. The GSM, which is conveniently formulated in the Rigged Hilbert Space,\cite{rf:12} offers a fully symmetric treatment of bound, resonance, and scattering single-particle states. The second scheme, the real-energy CSM,\cite{rf:3,rf:4} uses the projection formalism  in the Hilbert space. A recent realization of this approach is the Shell Model Embedded in the Continuum (SMEC)\cite{rf:13}, which offers a unified description of the structure and reactions with up to two nucleons in the scattering continuum using realistic SM Hamiltonians. In the following, we shall employ SMEC to illustrate the mixing of SM eigenstates induced by the coupling to the decay channel. 

The detailed description of SMEC can be found in recent reviews\cite{rf:3,rf:14}. The Hilbert space is divided into orthogonal subspaces ${\cal Q}_{\mu}$ ($\mu=0,1,\dots$)  corresponding to the different number of particles in the scattering continuum. An OQS description of `internal' dynamics, i.e., in ${\cal Q}_0$, includes couplings to the environment of decay channels and is given by the energy-dependent effective Hamiltonian.  In the approximation of one-nucleon scattering continuum, this effective Hamiltonian reads:
\begin{eqnarray}
{\cal H}(E)=H_0+H_1(E)=H_0+V_0^2h(E),
\label{eq1}
\end{eqnarray}
where $H_0$ is the CQS Hamiltonian (the SM Hamiltonian), $V_0$ is the continuum-coupling constant, $E$ is the scattering energy, and $h(E)$ is the  term describing the coupling  between localized states (${\cal Q}_0$) and the environment of one-nucleon decay channels ({\bf ${\cal Q}_1$}). The `external' mixing of two SM eigenstates $i$ and $j$ due to $H_1(E)$ consists of the Hermitian principal value integral describing virtual continuum excitations  and the anti-Hermitian residuum that represents the irreversible decay out of the internal space ${\cal Q}_0$. 

An explicit energy dependence of the effective Hamiltonian is the origin of its strong non-linearity. Moreover, the continuum-coupling term  generates effective many-body interactions in the internal space, even if it has  two-body character in the full space. Therefore, one should not separate a problem of many-body interactions in model spaces from a problem which of the two formulations of the many-body frameworks is being used: the CQS framework or the OQS framework. 

The biorthogonal eigenstates $|\Phi_i\rangle$ and  $|\Phi_{\bar i}\rangle$ of ${\cal H}$ are linear combinations of SM eigenstates $|\psi_i\rangle$:
\begin{eqnarray}
\label{transf}
|{\psi}_i\rangle \rightarrow |\Phi_j\rangle = {\sum}_{i}^{} b_{ji}|{\psi}_i\rangle,
\end{eqnarray}
where $[b_{ji}]$ is an orthogonal matrix. The left $|\Phi_i\rangle$ and right $|\Phi_{\bar i}\rangle$ eigenstates are related by the operation of  complex conjugation. Each eigenstate of  ${\cal H}(E)$ is coupled to states in neighboring nuclei via a network of reaction channels, either closed or open. Contribution of different reaction channels to the total continuum coupling is non-uniform and spans over a considerable range of excitation energies \cite{rf:15}. 
 
 In general, the continuum coupling lowers binding energies of near-threshold CQS eigenstates. The continuum-coupling correlation energy to the SM eigenstate $|\psi_i\rangle$,
\begin{eqnarray}
E_{{\rm corr};i}^{(\ell)}(E) = \langle \Phi_{\bar i}|{\cal H}-H_0|\Phi_i\rangle \simeq V_0^2\langle\Phi_{\bar i}|h(E) |\Phi_i\rangle, 
\label{eqcorr}
\end{eqnarray}
depends on the structure of the SM eigenstate and on the nature of decay channels involved. 

The correlation energy is peaked  at the threshold only for the coupling to a neutron decay channel  with $\ell=0$\cite{rf:16}. For higher $\ell$-values and/or for charged particles such as protons, deuterons, $\alpha$-particles, etc., the centrifugal barrier  and/or the Coulomb barrier shift the maximum of correlation energy  above the threshold.

The continuum coupling correlation energy has a characteristic form of an approximately symmetric peak centered either at the threshold for $\ell=0$ neutron decay channel, or above it in the continuum in other cases. The full width at half maximum $\sigma$ varies from case to case but a typical value is in the range: 2 MeV$\leq\sigma\leq$3 MeV. 

\begin{figure}[tb]
\begin{center}
\includegraphics[scale=0.8,angle=00,clip=true]{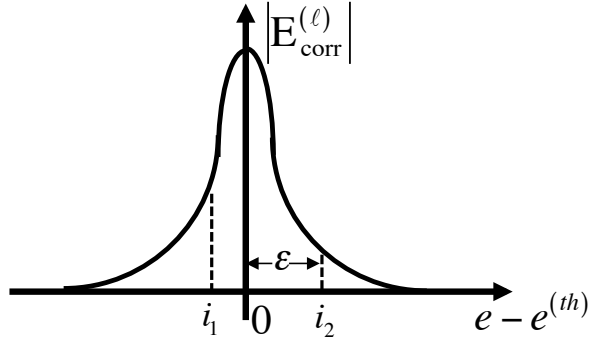}
\caption{\label{fig3}A schematic representation of the continuum-coupling correlation energy correction to the CQS eigenstates (SM eigenstates) as a function of the distance from the $\ell=0$ neutron decay threshold. The symbols  $i_1$ and  $i_2$ denote two eigenstates within the energy interval characteristic of strong coupling to the common decay threshold.}

\end{center}
\end{figure}

Figure \ref{fig3} shows a typical shape of the continuum-coupling correlation energy for SM states that are coupled to a neutron decay channel with $\ell=0$. The maximal value of $E_{{\rm (corr)};i}^{(\ell)}$ may vary from a few tenths of keV to several MeV, depending on the configuration of SM states involved and the nature of the decay channel. The relatively small value of the correlation energy is a consequence of the fact that only a few SM states of the same quantum numbers enter in the narrow window of strong correlations around the decay threshold. 

Even though the continuum-coupling correlation energy is small as compared to the binding energy, one should remember that it is of the same order as the pairing correlation energy, which profoundly modifies the independent particle motion in the atomic nucleus. 
It is therefore pertinent to ask whether correlations induced by the continuum coupling may lead to  instabilities of  certain SM eigenstates  near the channel threshold, and whether such couplings represent a collective phenomenon involving many SM eigenstates having the same quantum numbers. This question will be further addressed in Sec. \ref{sec4}.

For $E<0$ (bound system), the eigenvalues $E_i(E)$ of ${\cal H}(E)$ are real. In the continuum region, 
$E_i(E)$ correspond to the poles of the scattering matrix and ${\cal H}$ becomes complex-symmetric. The competition between Hermitian ($H_0$) and non-Hermitian ($H_1\equiv V_0^2h(E)$) parts of the effective Hamiltonian (\ref{eq1}) may lead to the coalescence of two eigenvalues, i.e., to the formation of the exceptional point\cite{rf:5}. Mathematically, EPs  are common roots of
\begin{eqnarray}
 \frac{\partial^{(\nu)}}{\partial {\cal E}} {\rm det}\left[{\cal H}\left(E;V_0\right)  -{\cal E}I\right] = 0,~~~{\rm with}~\nu=0, 1.
\label{discr}
\end{eqnarray}
Single-root solutions of Eq. (\ref{discr}) correspond to EPs associated with decaying (capturing) states. For the Hamiltonian (\ref{eq1}), the maximum number of such roots is:  
\begin{eqnarray}
M_{max}=2n(n-1),
\label{numroots}
\end{eqnarray}
where $n$ is the number of states of given angular momentum $J$ and parity $\pi$. The factor 2 in Eq.~(\ref{numroots}) comes from the quadratic dependence of ${\cal H}$ on $V_0$ (see Eq.~({\ref{eq1})), which yields identical solutions when  $V_0\rightarrow-V_0$. 

The wave function mixing in  CQS is closely related to features of exceptional points of the complex-extended CQS Hamiltonian. Similarly, to understand the configuration mixing in an OQS one must analyze the spectrum of EPs  of the complex-extended OQS Hamiltonian. 
Since the OQS Hamiltonian (\ref{eq1}) is energy dependent, the essential information about configuration mixing is contained in the trajectories of coalescing eigenvalues $E_{i_1}(E)=E_{i_2}(E)$  of the effective Hamiltonian -- the so-called exceptional threads -- for a complex value of the continuum coupling  $V_0$\cite{rf:17}. In the following section, we shall describe the mixing of SM eigenvalues close to the proton decay threshold in terms of the exceptional threads for the complex-extended SMEC Hamiltonian.

\section{Configuration mixing near the charge particle emission threshold}
\label{sec4}

Let us discuss the near-threshold configuration mixing and the role of exceptional threads  on the example of $J^{\pi}=0^+_i$, ($i=1,\dots,4)$ SM eigenstates of the two-proton emitter $^{16}$Ne \cite{Mukha} coupled to the $\ell=0$ proton decay channel leading to the  $1/2_1^{+}$  state of $^{15}$F. Our objective  is to understand generic aspects of the mixing between different realistic CQS eigenstates due to the coupling to the common $^{15}$F($1/2_1^+$)+p($\ell=0$)
decay channel rather than reproduce experimental data.

The SMEC calculations have been carried out  in the  $0p_{1/2}, 0d_{5/2}, 1s_{1/2}$ SM model space. In this  space, there are four   $J^{\pi}=0^+$ states in $^{16}$Ne. For $H_0$ we take the ZBM Hamiltonian \cite{rf:18}, which correctly describes the configuration mixing around the $N=Z=8$ shell closure. 
The residual coupling between ${\cal Q}_0$ and the embedding one-proton continuum ${\cal Q}_1$ is generated by the contact force $V_{12}=V_0\delta(r_1-r_2)$. We do not consider  the two-proton continuum   space ${\cal Q}_2$ so our model $^{16}$Ne is closed to two-proton decay.

The size of the continuum-coupling  correction to $H_0$ depends on two  parameters: the continuum coupling strength $V_0$ and the system excitation energy $E$ with respect to the threshold ($E=0$). The range of relevant $V_0$ values can be determined, for example, by fitting decay widths of the lowest states in $^{15}$F.  For a ZBM Hamiltonian, the experimental decay widths of the ground state $1/2_1^+$   and the first excited state $5/2_1^+$ in $^{15}$F are reproduced by taking  $V_0=-3500\pm 450$ MeV$\cdot$fm$^3$ and $V_0=-1100\pm 50$ MeV$\cdot$fm$^3$, respectively. The error bars in $V_0$ reflect experimental  width uncertainties. 

\subsection{Appearance of the aligned state}
\label{sec4.1}

Figure \ref{fig4} shows the predicted  energies and widths of the lowest $0^+$  states of   $^{16}$Ne. Experimentally, the proton separation energy is very small, $S_p \approx 100$ keV \cite{Mukha}, so all these states are  close to the one proton threshold (placed  at $E=0$ in Fig. \ref{fig4}). 
Below the proton threshold, the continuum-coupling correction is Hermitian and the eigenvalues of ${\cal H}$  are real. 
\begin{figure}[htb]
\begin{center}
\includegraphics[width=0.80\textwidth]{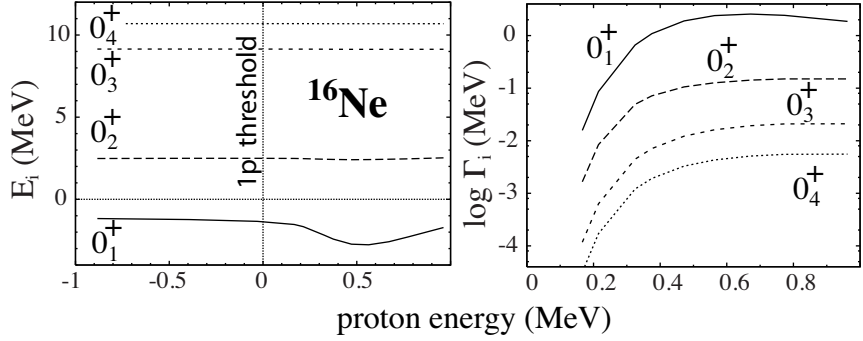}
\caption{\label{fig4}Energies $E_i$ (left) and widths $\Gamma_i$ (right) of the four lowest  $J^{\pi}=0^+$  eigenvalues of the effective Hamiltonian (\ref{eq1}) for  $^{16}{\rm Ne}$  as  a function of the proton energy $E$. The proton threshold at  $E=0$ is indicated by a vertical dotted line. The continuum coupling strength is $V_0=-1000~$MeV~fm$^3$.}
\end{center} 
\end{figure}
The lowest energy eigenvalue shows pronounced dependence on the continuum coupling in the vicinity of the threshold, as seen in Fig.~\ref{fig5}. The minimum of $E_1$ is shifted above the threshold by about 0.55 MeV  due to the Coulomb interaction. The width $\Gamma_i$ of all eigenvalues becomes different from zero only above the  threshold.  The maximum of $\Gamma_1(E)$  is found close to the energy corresponding to the minimum of the real part $E_1(E)$. At higher energies, $\Gamma_1(E)$ gradually  decreases\cite{rf:3}.
As seen in Figs.~\ref{fig4} and \ref{fig5},  continuum coupling has the strongest effect on the lowest-energy eigenvalue with the largest width and the correlation energy rapidly decreases for excited states.
\begin{figure}[htb]
\begin{center}
\includegraphics[width=0.50\textwidth]{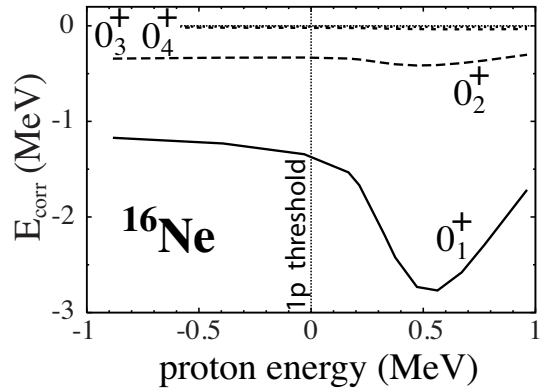}
\caption{\label{fig5}Similar as in Fig.~\ref{fig4} except for the continuum-coupling correlation energy (\ref{eqcorr}).}
\end{center} 
\end{figure}
\begin{figure}[htb]
\begin{center}
\includegraphics[width=0.70\textwidth]{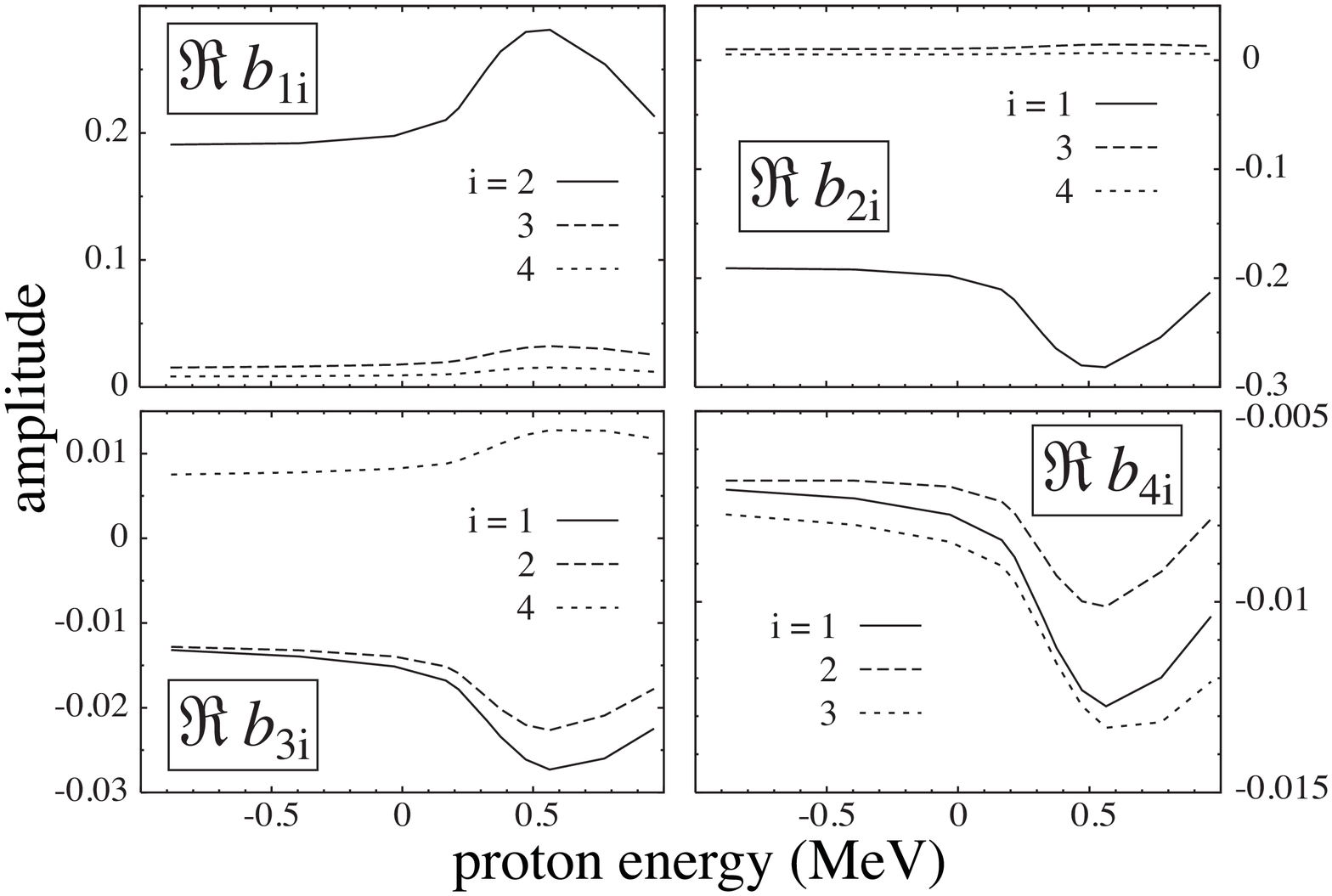}
\end{center} 
\caption{\label{fig6}The real part of the coefficients $b_{ji}$ of the orthogonal transformation (\ref{transf}) for $0_j^+$ eigenstates in $^{16}$Ne.}
\end{figure}
Figure \ref{fig6} shows the real part of the off-diagonal matrix elements of the orthogonal transformation $[b_{ji}]$ (Eq. (\ref{transf})) corresponding  to the coupling of $0_j^+$, ($j=1,\dots, 4$) SM eigenstates in $^{16}$Ne to the one-proton $\ell=0$ decay channel. The maximum/minimum of $b_{ji}(E)$ indicates the region of the strongest mixing at the maximum of the continuum-coupling correlation energy ($E_{max}\simeq0.55$ MeV). (The imaginary part of $b_{ji}(E)$ exhibits similar behavior.) The large matrix elements involve the  $0_1^+$ state of ${\cal H}$, which couples strongest to the decay threshold. This single state,
which we shall call the 
`aligned state' in the following,  couples strongest to the decay channel; hence, it carries many of its characteristics. Aligned states 
are crucial for understanding the mixing of SM eigenstates around the reaction threshold.

\subsection{Exceptional threads}
\label{sec4.2}
\begin{figure}[htb]
\begin{center}
\includegraphics[width=0.60\textwidth]{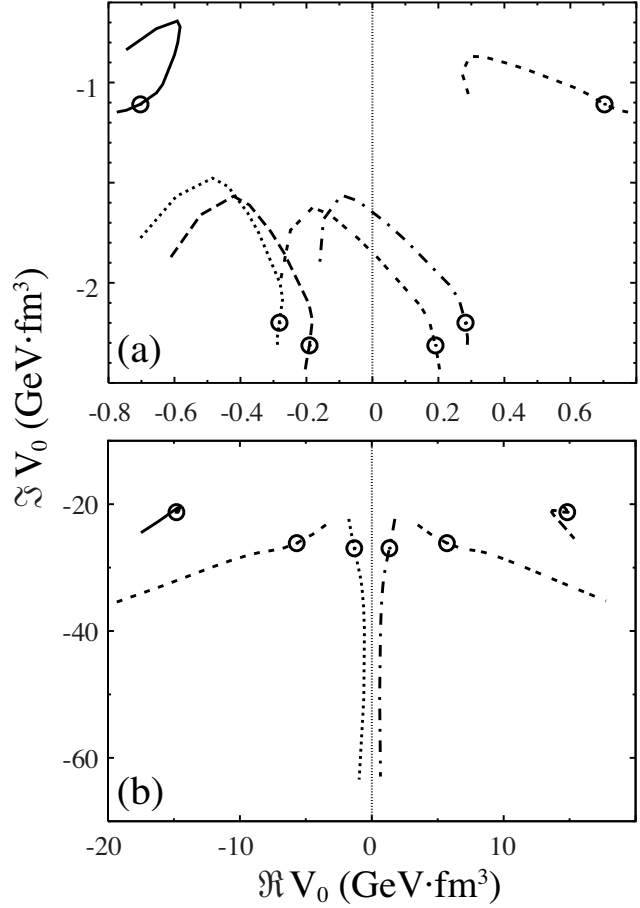}
\caption{\label{fig7}Exceptional threads for the $0^+$ eigenvalues of the SMEC Hamiltonian for $^{16}$Ne   in the complex-$V_0$ plane. Each point at the exceptional thread corresponds to an exceptional point at a definite energy $E$. The  threshold position ($E=0$) is indicated for each thread with a circle. Because of the invariance of ${\cal H}$ to $V_0\rightarrow-V_0$ only half of all possible exceptional threads are shown in the region of small (a) and very large (b)  values of $V_0$.
\newline
 Exceptional threads in (a) formed by the $0_1^+$-$0_2^+$ coalescence are marked by solid  and short-dashed lines for decaying and capturing double-poles, respectively, while 
the ($0_1^+$, $0_4^+$) and ($0_1^+$, $0_3^+$) pairs are indicated by
long dashed and double-short-dashed, and dotted and dashed-dotted lines, respectively. 
 Exceptional threads in (b) do not involve the aligned state $0_1^+$.}
\end{center}
\end{figure}
A complete picture of the near-threshold configuration mixing in CSM many-body wave functions can be obtained by investigating properties of exceptional threads of the complex-extended CSM Hamiltonian. Figure \ref{fig7} shows the exceptional threads in the complex-$V_0$ plane for the  $0^+$ states of  $^{16}$Ne.   According to (\ref{numroots}) there are 12 doubly-degenerate exceptional points for 4 eigenstates. Six of these correspond to pairs of eigenvalues with the outgoing asymptotics (decaying resonances) and six to pairs of eigenvalues with the incoming asymptotics (capturing resonances). For $E<0$, the effective Hamiltonian (\ref{eq1}) is Hermitian; hence, exceptional threads for capturing and decaying resonances have identical $\Im V_0$ and their real parts have opposite signs. In general, threads for $\Re V_0 \cdot  \Im V_0 > 0$ ($\Re V_0 \cdot  \Im V_0 < 0$) correspond to decaying (capturing) states. This symmetry does not hold for $E>0$.

Figure \ref{fig7}(a) shows six threads in the region of small values of the coupling constant. Each of those double poles is  a result of the coalescence of the lowest-energy eigenvalue $0_1^+$ of the effective Hamiltonian with one of the remaining three eigenvalues. 
All eigenstates are mutually mixed  through their couplings with the aligned state $0_1^+$. All exceptional threads relevant  to   the mixing involve this aligned eigenstate. As seen in Fig. \ref{fig5},  the aligned eigenstate exhausts $\sim 80\%$ of the continuum-coupling correlation energy. This is a generic feature of aligned states, which we also found in other systems close to the $\ell=0$ proton decay threshold. 

One should notice that the point that is closest to the physical limit of the SMEC Hamiltonian ($\Im V_0=0$), the turning point, has moved away from the threshold into the scattering continuum. 
This turning point of all exceptional threads involving the aligned state appears consistently at the same energy $E_{max}$ that corresponds to a minimum of  $E_1$ (see Fig. \ref{fig4}), a maximum of $E_{corr}$ in $0_1^+$ (see Fig. \ref{fig5}), and extrema of $b_{1k}(E)$ (see Fig. \ref{fig6}). 

The remaining six double-poles of the scattering matrix are found at exceedingly large values of  complex $V_0$, outside the range of physically significant values, as seen in  Fig. \ref{fig7}(b). In this case,  all exceptional threads for capturing resonances remain in the $\Re V_0>0$ half-plane  also for $E>0$. With increasing energy of the system, one pair of double poles escapes towards $\Re V_0=\pm\infty$, the second pair escapes towards $\Im V_0=\pm\infty$, and the third one remains confined in a relatively narrow range of $V_0$.

\subsection{Maximum continuum coupling point}
\label{sec4.3}

As discussed above, the continuum-mixing among $0_i^+$ SM eigenstates of $^{16}$Ne  is strongest at the turning point at $E_{max}\simeq 0.55$\,MeV above the $\ell=0$ proton-decay threshold. 
In general, the scattering energy  corresponding to the turning point depends on the height of the effective (Coulomb+centrifugal)  barrier, structure of SM wave functions, and the channel itself. In the studied case of $\ell=0$ decay, the centrifugal barrier is absent. The height of the Coulomb barrier for $Z_t=Z-1=9$ in $^{16}$Ne is $V_C=1.69\pm0.01$ MeV. This estimate has been obtained by considering  a uniformly charged sphere of radius $R_0=1.27(A-1)^{1/3}$ fm. The error bar on $V_C$ corresponds to the variation $U_0=52\pm5$ MeV of the depth of the Woods-Saxon (WS) potential of radius $R_0$ and diffuseness  $d=0.67$ fm. 

To see the dependence of $E_{max}$ on $V_C$, we investigate  the mixing of two $0^+$ SM states in $^{24}$S due to the coupling to the one-proton decay channel
$^{23}$P($1/2_1^+$)+p($\ell=0$).
In this case, there are two exceptional threads associated with decaying and capturing resonances. SMEC calculations are performed in the $1s_{1/2}, 0d_{5/2}, 0d_{3/2}$ SM model space using the USDB interaction\cite{rf:19}. 
\begin{figure}[htb]
\begin{center}
\includegraphics[width=0.60\textwidth]{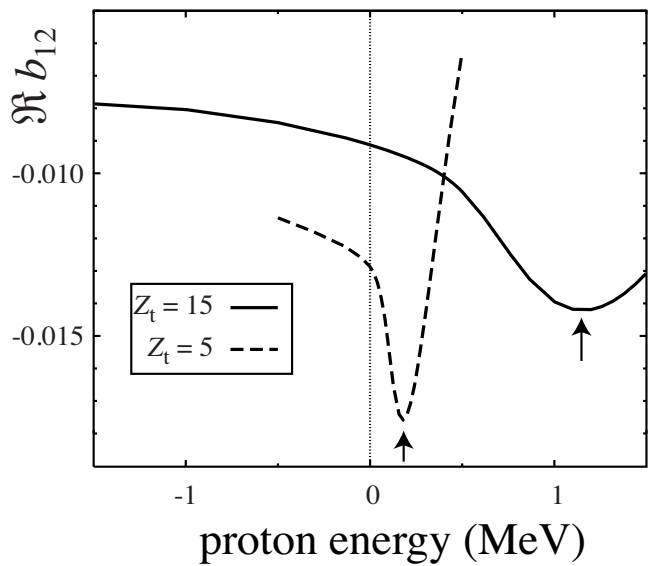}
\caption{\label{fig8}The real part of $b_{12}$ in $^{24}$S ($Z_t=15$, solid line). The dashed line shows results with  the Coulomb potential reduced by a factor of three ($Z_t=5$). The turning points are marked by arrows.}
\label{fig8}
\end{center}
\end{figure}

Figure \ref{fig8} shows the energy dependence of the real part of the mixing matrix element $b_{12}(E)$. For $^{24}$S, the height of the Coulomb barrier is $V_C=2.76\pm0.03$ MeV. To assess the impact of the Coulomb barrier, we also carried out calculations with the reduced Coulomb potential having  $V_C=0.82\pm0.007$ ($Z_t=5$). One can see that $E_{max}$  strongly depends  on $V_C$.  
The turning point of exceptional threads in $^{24}$S  is 1.25 MeV and 0.17 MeV for standard  and reduced Coulomb potentials, respectively. Notice that the magnitude of mixing, as given by the amplitude  $|b_{12}(E)|$, is strongly reduced with increasing charge $Z_t$. This means that the imprint of the branching point at a charge particle decay threshold on  regular solutions of a scattering problem is shielded by the Coulomb barrier; hence, it is reduced in heavier nuclei. This generic feature, with far-reaching consequences for the manifestation of nuclear clustering, does not depend on the nature of the decay channel.

\begin{figure}[htb]
\begin{center}
\includegraphics[width=0.70\textwidth]{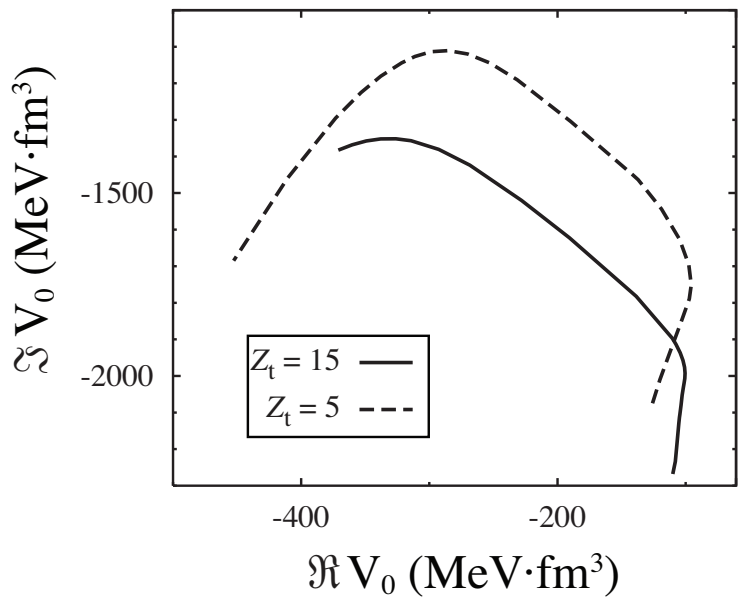}
\caption{\label{fig9}Exceptional threads for the $0^+$ states in  $^{24}$S  in the 
complex-$V_0$ plane for the standard  (solid line) and reduced (dashed line)  Coulomb barrier.}
\end{center} 
\end{figure}
Finally, Fig.~\ref{fig9} shows the exceptional threads in the complex-$V_0$ plane for the $0^+$ states in  $^{24}$S.  We show here only those exceptional threads which are associated with decaying resonances.
It is seen that
the turning point moves further away from the physical limit at $\Im V_0$=0
with increasing charge; this nicely illustrates the findings of Fig.~\ref{fig8}, i.e., the continuum-coupling induced configuration mixing diminishes with $Z_t$.
This generic effect due to the Coulomb barrier not only suppresses proton halos in heavier nuclei, but also prevents $\alpha$ clustering  near $\alpha$ emission thresholds in heavy systems. Conventionally, the latter effect has been associated with the disruptive influence of the nuclear spin-orbit interaction on $\alpha$-clustering (see Ref.\cite{rf:20} and references cited therein), i.e., with a specific feature of the nuclear potential.

\section{Conclusions and perspectives}
\label{sec5}
The phenomenon of nuclear clustering has been a key problem  in low-energy nuclear structure studies. 
In this work, we have applied  the OQS framework of SMEC to study the behavior  of SM eigenstates in the vicinity of the $\ell=0$ proton-decay threshold.  Proximity of the branching point induces the collective mixing of  SM eigenstates, in which the essential role is played by a single  aligned eigenstate of the OQS Hamiltonian. The aligned state, in which most of  the continuum-coupling correlation energy is concentrated, is present in all exceptional threads close to the physical limit of the OQS Hamiltonian ($\Im V_0$=0) for realistic values of the real part of the coupling constant $\Re V_0$.  Above the decay threshold, this state becomes a broad resonance. 

The collective mixing of SM eigenstates via the aligned state -- an imprint of the decay channel  --  is the essence  of the clustering phenomenon. In the studied examples of the coupling to $\ell=0$ proton decay channel, an aligned  state  can cause an instability of a  near-threshold SM eigenstate if its energy is close to the turning point of associated  exceptional threads. We thus conjecture that clustering present  in the vicinity of {\em any} charged-particle emission threshold is a consequence of a mixing mediated by an aligned state involving  {\em all} SM eigenstates with the same quantum numbers $J^\pi$. Hence, the presence of charged-cluster states near the respective charged-cluster emission threshold is  a signature of a profound change of the near-threshold SM wave function and the direct  manifestation of  continuum-coupling correlations. 

In order to see an imprint of the aligned state in a near-threshold SM wave function, the distance of the turning point from a  real axis $\Im V_0$=0 should be small. This distance, controlled mainly by the Coulomb barrier penetrability, increases rapidly with increasing proton number. This feature, which is generic for all configuration mixing situations involving the $\ell=0$ charged-particle (e.g. proton, deuteron, $\alpha$-particle) emission, reduces clustering correlations in heavier nuclei. 

Favorable conditions for the appearance of charged-cluster configurations are {\em above} the charged-cluster emission threshold. On the other hand,  neutral-cluster configurations are expected to show up primarily {\em below} the threshold -- due to the rapid growth  of the decay width with energy. Here, spectacular examples are one- and two-neutron halos in light nuclei. 

In summary, we have shown that the many-body correlations due to continuum coupling  can profoundly impact the nature of states near reaction thresholds.  While much work is still needed to fully understand the phenomenon of nuclear clustering, the recent progress in microscopic nuclear theory of OQSs, and  insights 
provided by CSM, make us optimistic about the solution to this fascinating puzzle.

\section*{Acknowledgements}
This work has been supported in part by the MNiSW grant No. N N202 033837; the Collaboration COPIN-GANIL; and  U.S. Department of Energy under Contract Nos.
DE-FG02-96ER40963 (University of Tennessee) and DE-FG02-10ER41700 (French-U.S. Theory Institute for Physics with Exotic Nuclei).


\begin{thebibliography}{99}
\bibitem{rf:1} A.~Bohr and B.R.~Mottelson, {\em Nuclear Structure} (Benjamin, New York, 1975), Vol II;\\
O.~Haxel, J.H.D.~Jensen and H.E.~Suess, Phys.\ Rev.\ \textbf{75} (1949), 1766;\\
M.G.~Mayer, Phys.\ Rev.\ \textbf{75} (1949), 1969;\\
A.M.~Lane, Proc.\ R.\ Soc.\ \textbf{A 68} (1955), 197;\\
D.~Kurath, Phys.\ Rev.\ \textbf{101} (1956), 216;\\
E.~Caurier, G.~Martinez-Pinedo, F.~Nowacki, A.~Poves and A.P.~Zuker, Rev.\ Mod.\ Phys.\ \textbf{77} (2005), 427.
\bibitem{rf:2} 
J.A.~Wheeler, Phys.\ Rev.\ \textbf{52} (1937), 1083;\\
C.F.~von~Weizs\"acker, Naturwiss.\ \textbf{26} (1938), 209;\\
W.~Wefelmeier, Z.\ Phys.\ \textbf{107} (1937), 332.
\bibitem{rf:3} 
J.~Oko{\l}owicz, M.~P{\l}oszajczak and I.~Rotter, Phys.\ Rep.\ \textbf{374}, (2003), 271.
\bibitem{rf:4} 
A.~Volya and V.~Zelevinsky, Phys.\ Rev.\ C\textbf{74} (2006), 064314.            
\bibitem{rf:5}  T.~Kato, {\em Perturbation Theory for Linear Operators} (Springer Verlag, Berlin, 1995);\\
M.R.~Zirnbauer, J.J.M.~Verbaarschot and H.A.~Weidenm\"{u}ller, Nucl.\ Phys.\  \textbf{A411} (1983), 161;\\
W.D.~Heiss and W.-H.~Steeb, J.\ Math.\ Phys.\ \textbf{32} (1991), 3003.
\bibitem{rf:6} K.~Ikeda, N.~Takigawa and H.~Horiuchi, Prog.\ Theor.\ Phys.\ Suppl\ Extra number (1968), 464.
\bibitem{rf:7} A.I.~Baz, Ya.B.~Zel'dovich and A.M.~Perelomov, {\em Scattering Reactions and Decay 
in Nonrelativistic Quantum Mechanics}, Israel\ Program\ for\ Scientific\ Translations\, Jerusalem (1969).
\bibitem{rf:8} N.~Michel, W.~Nazarewicz and M.~P{\l}oszajczak, Phys.\ Rev.\ C\textbf{75} (2007), 031301.
\bibitem{rf:9} N.~Michel, W.~Nazarewicz and M.~P{\l}oszajczak, Nucl.\ Phys.\ \textbf{A794} (2007), 29.
\bibitem{rf:10} N.~Michel, W.~Nazarewicz, M.~P{\l}oszajczak and K.~Bennaceur, Phys.\ Rev.\ Lett.\   \textbf{89} (2002), 042502;\\
R.~Id Betan, R.J.~Liotta, N.~Sandulescu and T.~Vertse, Phys.\ Rev.\ Lett.\ \textbf{89} (2002), 042501;\\
N.~Michel, W.~Nazarewicz, M.~P{\l}oszajczak  and Oko{\l}owicz, Phys.\ Rev.\  C\textbf{67} (2003), 054311.
\bibitem{rf:11} T.~Berggren, Nucl.\ Phys.\  A \textbf{109} (1968), 265.
\bibitem{rf:12} I.M.~Gel'fand and N.Ya.~Vilenkin,  {\it Generalized Functions, Vol. 4} (Academic Press, New York) (1961);\\
K.~Maurin, {\it Generalized Eigenfunction Expansions and Unitary Representations of Topological Groups} (Polish Scientific Publishers, Warsaw) (1968).
\bibitem{rf:13} K.~Bennaceur, F.~Nowacki, J.~Oko{\l}owicz and M.~P{\l}oszajczak, Nucl.\ Phys.\ A \textbf{651} (1999), 289;\\
J.~Rotureau, J.~Oko{\l}owicz and M.~P{\l}oszajczak, Nucl.\ Phys.\   A \textbf{767} (2006), 13.
\bibitem{rf:14} B.~Blank and M.~P{\l}oszajczak, Rep.\ Prog.\ Phys.\ \textbf{71} (2008), 046301. 
\bibitem{rf:15} J.~Oko{\l}owicz, M.~P{\l}oszajczak and Y.~Luo, Acta\ Phys.\ Pol.\ B \textbf{39} (2007), 389.
\bibitem{rf:16} J.~Oko{\l}owicz and M.~P{\l}oszajczak, (2012), to be published.
\bibitem{rf:17} J.~Oko{\l}owicz and M.~P{\l}oszajczak, Acta\ Phys.\ Pol.\ B \textbf{42} (2011), 451.
\bibitem{Mukha} I. Mukha {\it et al.}, Phys.\ Rev.\ C \textbf{82} (2010), 054315.
\bibitem{rf:18} A.P.~Zuker, B.~Buck and J.B.~McGrory, Phys.\ Rev.\ Lett.\ \textbf{21} (1968), 39.
\bibitem{rf:19} B.A.~Brown and W.A.~Richter, Phys.\ Rev.\ C \textbf{74} (2006),  034315.
\bibitem{rf:20} N.~Itagaki, J.~Cseh and M.~P{\l}oszajczak, Phys.\ Rev.\ C {\bf 83} (2011), 044315.
\end{thebibliography}
\end{document}